\documentclass{ws-ijmpe}
\usepackage{graphicx}
\usepackage{amsmath}

\newcommand{\beq}{\begin{equation}}
\newcommand{\eeq}{\end{equation}}
\newcommand{\beqa}{\begin{eqnarray}}
\newcommand{\eeqa}{\end{eqnarray}}
\newcommand{\bseq}{\begin{subequations}}
\newcommand{\eseq}{\end{subequations}}

\def\x{{\boldsymbol x}}

\def\p{{\boldsymbol p}}
\def\0{{\boldsymbol 0}}

\def\kk{{\kappa}}
\def\pp{{\hat{p}}}

\def\simle{\mathrel{\rlap{\raise 0.511ex \hbox{$<$}}{\lower 0.511ex 
\hbox{$\sim$}}}}
\def\simge{\mathrel{ \rlap{\raise 0.511ex 
\hbox{$>$}}{\lower 0.511ex \hbox{$\sim$}}}}

\begin{document}

\title{Langevin dynamics of heavy flavors in relativistic heavy-ion collisions}

\author{\footnotesize W.M. ALBERICO$^{3,4}$, A. BERAUDO$^{1,2,3,4}$, A. DE PACE$^4$, A. MOLINARI$^{3,4}$, M. MONTENO$^4$, M. NARDI$^4$  and F. PRINO$^4$}
\address{
$^1$Centro Studi e Ricerche \emph{Enrico Fermi}, Piazza del Viminale 1, Roma, ITALY\\
$^2$Physics Department, Theory Unit, CERN, CH-1211 Gen\`eve 23, Switzerland\\
$^3$Dipartimento di Fisica Teorica, Universit\`a di Torino, via P. Giuria 1, I-10125 Torino, Italy\\
$^4$Istituto Nazionale di Fisica Nucleare, Sezione di Torino, via P.Giuria 1, I-10125 Torino, Italy\\
alberico@to.infn.it, beraudo@to.infn.it, depace@to.infn.it, molinari@to.infn.it, monteno@to.infn.it, nardi@to.infn.it {\rm and} prino@to.infn.it}

\maketitle

\begin{abstract}
We study the stochastic dynamics of $c$ and $b$ quarks, produced in hard initial processes, in the hot medium created after the collision of two relativistic heavy ions. This is done through the numerical solution of the relativistic Langevin equation. The latter requires the knowledge of the friction and diffusion coefficients, whose microscopic evaluation is performed treating separately the contribution of soft and hard collisions. The evolution of the background medium is described by ideal/viscous hydrodynamics. Below the critical temperature the heavy quarks are converted into hadrons, whose semileptonic decays provide single-electron spectra to be compared with the current experimental data measured at RHIC. We focus on the nuclear modification factor $R_{AA}$ and on the elliptic-flow coefficient $v_2$, getting, for sufficiently large $p_T$, a reasonable agreement.
\end{abstract}

\section{Introduction}
Heavy quarks, produced in initial hard processes, allow to perform a ``tomography'' of the medium (hopefully a Quark Gluon Plasma) created in heavy-ion collisions. The modification of their spectra provides information on the properties (encoded into few transport coefficients) of the matter crossed before hadronizing and giving rise to experimental signals (so far, the electrons from their semi-leptonic decays). We employ an approach based on the relativistic Langevin equation, assuming that medium-modifications of the initial heavy-quark spectrum (in particular its quenching at high $p_T$) arise from the cumulative effect of many random collisions. The final goal of our analysis is to achieve a reasonable description of the $R_{AA}$ and $v_2$ of non-photonic electrons from heavy-flavor decays measured at RHIC. Preliminary results can be found in Ref.~\cite{lange_hot}. For similar studies see also Refs.~\cite{hira,rapp,tea,aic}.

\section{The Langevin equation in a dynamical medium}
We solve the Langevin equation for heavy quarks propagating in the expanding fireball created in heavy-ion collisions. We assume that local thermal equilibrium is reached, so that the background medium is entirely described by the four-velocity and temperature fields $u^\mu(x)$ and $T(x)$, provided by two different hydro codes~\cite{kolb1,rom1,rom2}. This requires a generalization of the algorithm used in Ref.~\cite{lange} for the static case. 
We focus on a given quark at $(\x_n,\p_n)$ after $n$ steps of evolution. Next we move to the local fluid rest-frame and update its position and momentum by the quantities $\Delta\bar{\x}_n\!=\!(\bar{\p}_n/\bar E_p)\Delta\bar t$ and
\beq
\Delta\bar{p}_n^i=-\eta_D(\bar{p}_n)\bar{p}_n^i\Delta\bar t+\xi^i(\bar t)\Delta\bar{t}
\equiv-\eta_D(\bar p_n)\bar p_n^i\Delta \bar t+{g^{ij}(\bar\p_n)}{\zeta^j(\bar t)}\sqrt{\Delta\bar t},
\eeq
where we take -- in the fluid rest-frame -- $\Delta\bar t\!=\!0.02$ fm/c. In the above we express the noise term through the tensor (we omit the ``bar'')
\beq
{g^{ij}(\p)\!\equiv\!\sqrt{\kk_L(p)}\pp^i\pp^j+\sqrt{\kk_T(p)}
(\delta^{ij}-\pp^i\pp^j)},
\eeq
depending on the transverse/longitudinal momentum diffusion coefficients $\kappa_{T/L}(p)$, and the uncorrelated random variables $\zeta^j$, with $\langle\zeta^i(t)\zeta^j(t')\rangle\!=\!\delta^{ij}\delta_{tt'}$.
Hence, one simply needs to extract three random numbers $\zeta^j$ from a gaussian distribution with $\sigma\!=\!1$.
Then, going back to the Lab frame, one obtains the updated $(\x_{n+1},\p_{n+1})$. 

\section{Evaluation of the transport coefficients}
\begin{figure}
\begin{center}
\includegraphics[clip,width=0.56\textwidth]{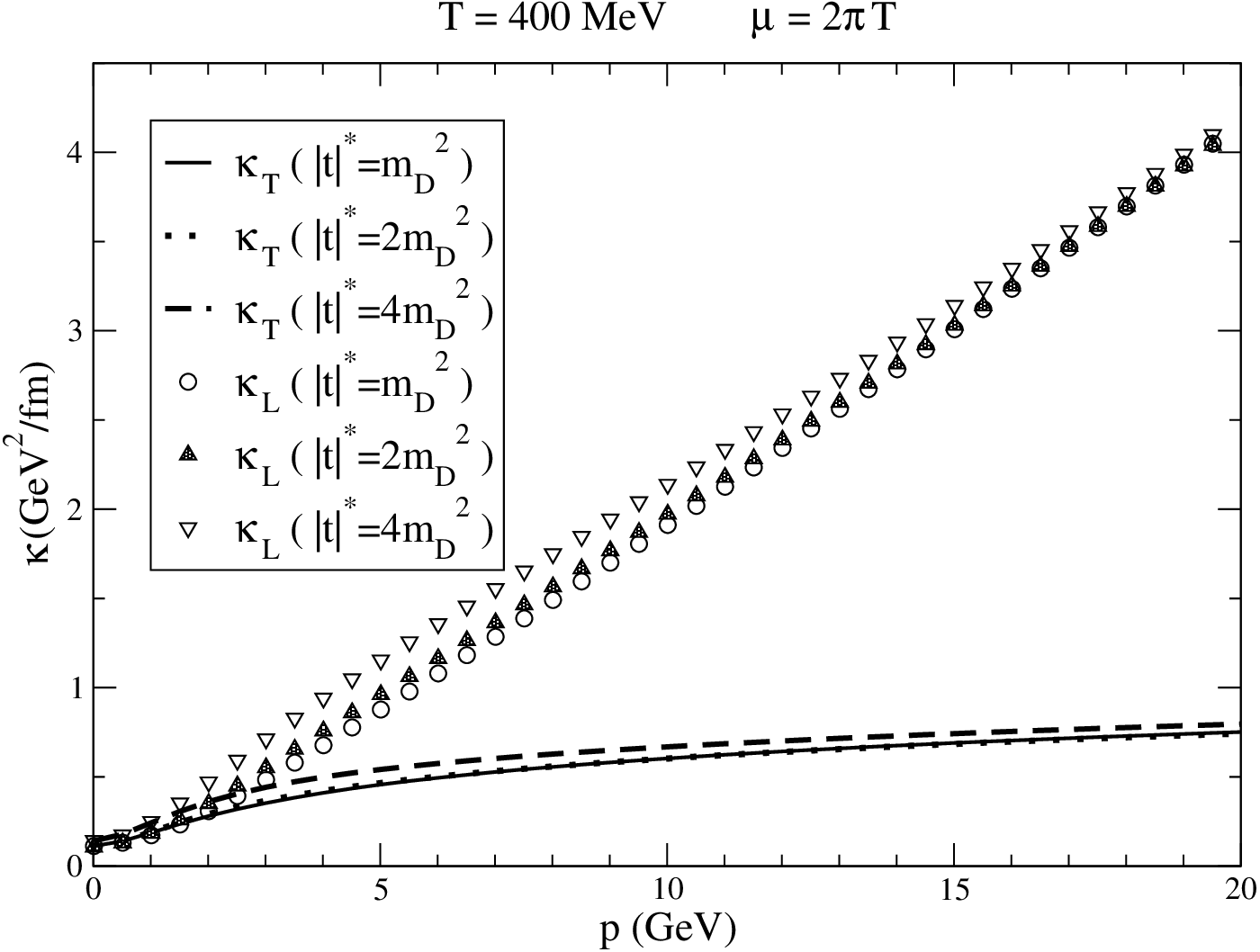}
\caption{The momentum-diffusion coefficients $\kappa_{T/L}$ of a $c$ quark after summing the soft and hard contributions. The dependence on the cutoff $|t|^*$ is mild. The coupling $g$ was evaluated at $\mu\!=\!2\pi T$.} 
\label{fig:transport}
\end{center}
\end{figure}

The coefficients
\beq
\kappa_T(p)\equiv\frac{1}{2}\frac{\langle \Delta p_T^2\rangle}{\Delta t}\quad{\rm and}\quad\kappa_L(p)\equiv\frac{\langle \Delta p_L^2\rangle}{\Delta t}
\eeq
yield the average transverse/longitudinal squared-momentum acquired per unit time by the heavy quark through the collisions in the medium. Following Ref.~\cite{pei} we introduce an intermediate cutoff $|t|^*\!\sim\!m_D^2$ ($t\!\equiv\!(P'\!-\!P)^2$) separating hard and soft scatterings. The contribution of hard collisions ($|t|\!>\!|t|^*$) is evaluated through a kinetic pQCD calculation of the processes $Q(P)q_{i/\bar i}\!\to\! Q(P')q_{i/\bar i}$ and $Q(P)g\!\to\! Q(P')g$. On the other hand in soft collisions ($|t|\!<\!|t|^*$) the exchanged gluon has ``time'' to feel the presence of the other particles. A resummation of medium effects is thus required and this is provided by the Hard Thermal Loop approximation. The final result is given by the sum of the two contributions $\kappa_{T/L}(p)=\kappa_{T/L}^{\rm hard}(p)+\kappa_{T/L}^{\rm soft}(p)$ and its explicit expression can be found in Ref.~\cite{lange_hot}. In Fig.~\ref{fig:transport} we display its behavior in the case of a $c$ quark in a plasma at $T\!=\!400$ MeV. The sensitivity to the value of the intermediate cutoff $|t|^*$ is quite small, hence supporting the validity of the approach. 
Finally, as discussed in Ref.~\cite{lange}, the friction coefficient $\eta_D(p)$ is fixed so that the heavy quarks evolve towards thermal equilibrium. 
\section{Numerical results}
\begin{table}
\begin{center}
\begin{tabular}{|c|c|c|}
\hline
{}& $\sigma_{c\bar c}\,(\mu b)$ & {$\sigma_{b\bar b}\,(\mu b)$}\\
\hline
pp & 254.14 & 1.769\\
\hline
AA & 236.11 & 2.033\\
\hline
\end{tabular}
\begin{tabular}{|c|c|c|c|}
\hline
Hydro code & $\tau_0$ (fm/c) & $s_0$ (fm$^{-3}$) & $T_0$ (MeV)\\
\hline
ideal & 0.6 & 110 & 357\\
\hline
viscous & 1.0 & 83.8 & 333\\
\hline
\end{tabular}
\caption{Initialization for RHIC: the $c\bar c$ and $b\bar b$ production cross section at $\sqrt{s}_{NN}\!=\!200$ GeV given by POWHEG and the explored hydro scenarios.}\label{table:cross_hydro}
\end{center}
\end{table}
For each explored case we generated an initial sample of $45\!\cdot\!10^6$ $c\bar c$ and $b\bar b$ pairs, using the POWHEG code~\cite{POWHEG}, with CTEQ6M PDFs. In the $AA$ case we introduced nuclear effects in the PDFs according to the EPS09 scheme~\cite{EPS09}; the quarks were then distributed in the transverse plane according to the nuclear overlap function ${dN/d\x_\perp\!\sim\!T_{AB}(x,y)}\!\equiv\!T_A(x\!+\!b/2,y)T_B(x\!-\!b/2,y)$ and given a further $k_T$ broadening on top of the ``intrinsic'' one.
At the proper-time $\tau\!\equiv{t^2\!-\!z^2}\!=\!\tau_0$ we started following the Langevin dynamics of the quarks until hadronization. The latter was modeled using Peterson fragmentation functions~\cite{peter}, with branching fractions into the different hadrons taken from Refs.~\cite{zeus,pdg}.
Finally each hadron was forced to decay into electrons with PYTHIA~\cite{Pythia}, using updated decay tables~\cite{pdg09}.
The $e$-spectra from $c$ and $b$ were then combined with a weight given by the respective total production cross-section quoted in Table~\ref{table:cross_hydro}, where the hydro initial conditions are also reported.
\begin{figure}
\begin{center}
\includegraphics[clip,width=0.495\textwidth]{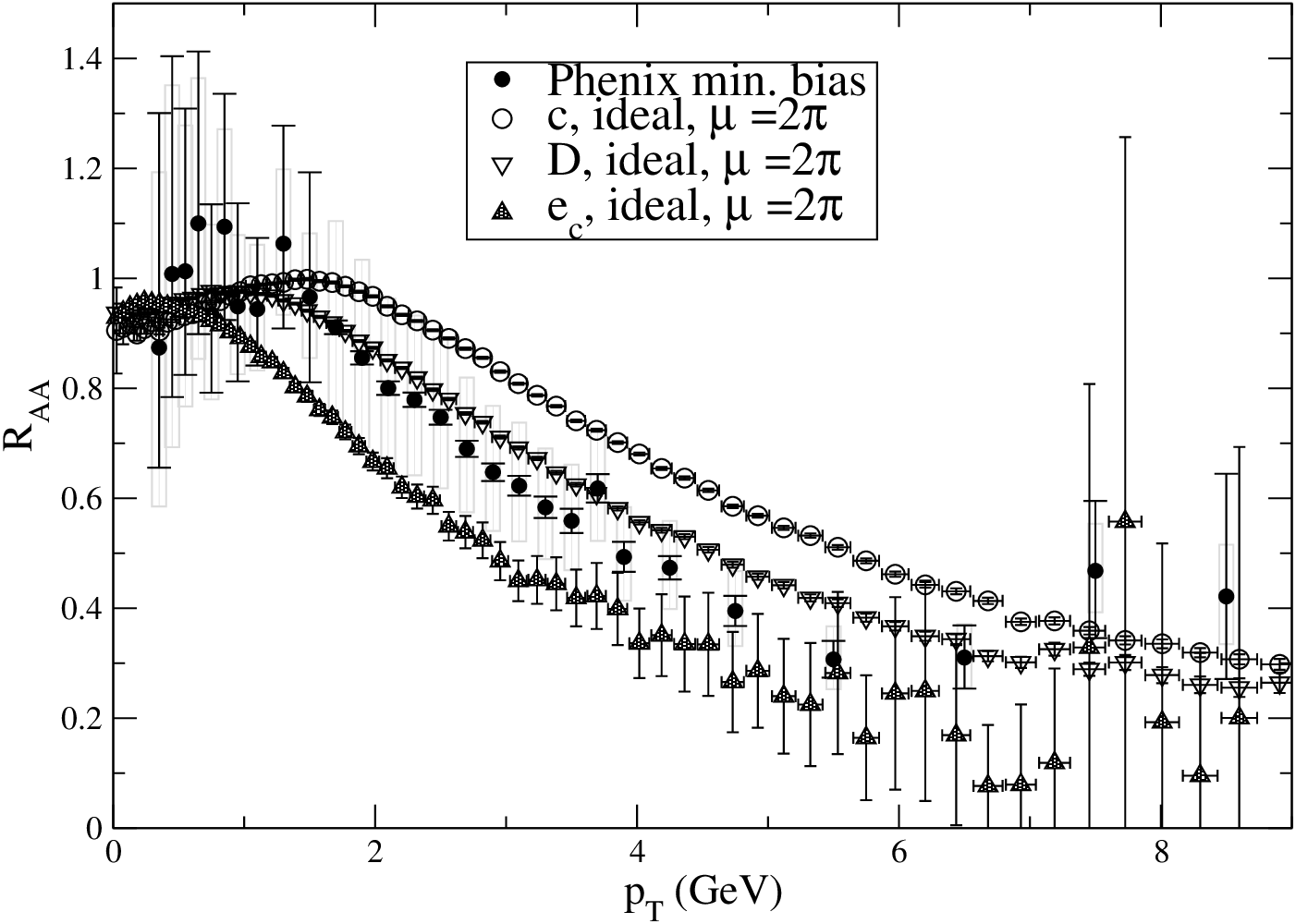}
\includegraphics[clip,width=0.495\textwidth]{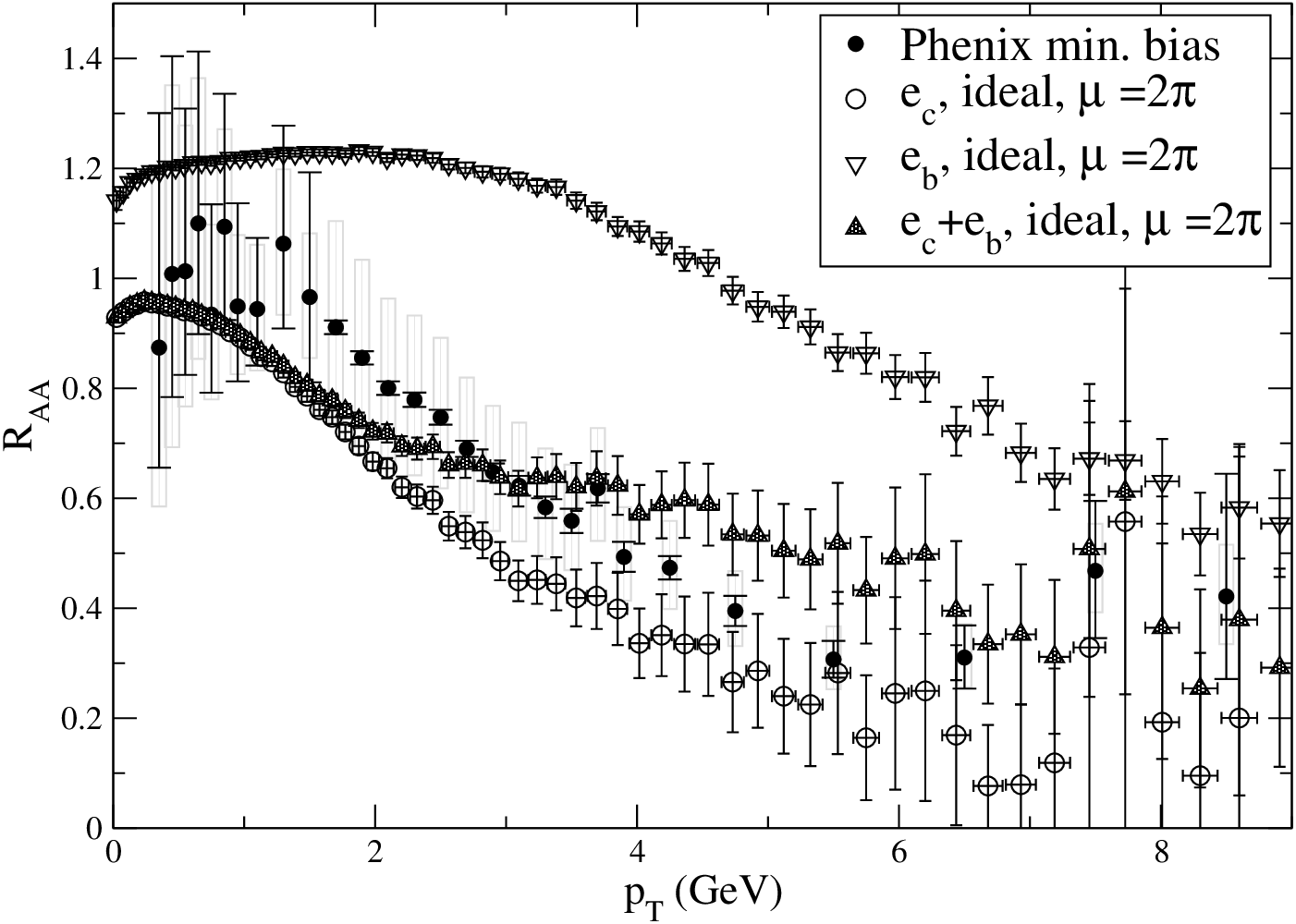}
\caption{Left panel: the effect of fragmentation and semi-leptonic decays on the charm $R_{AA}$, both leading to a quenching of the spectrum. Right panel: the separate contribution of charm and bottom to the single-electron $R_{AA}$. Electrons from $b$-decays start contributing for $p_T\!\simge\!2$ GeV/c.} 
\label{fig:RAA_system}
\end{center}
\end{figure}
\begin{figure}
\begin{center}
\includegraphics[clip,width=0.495\textwidth]{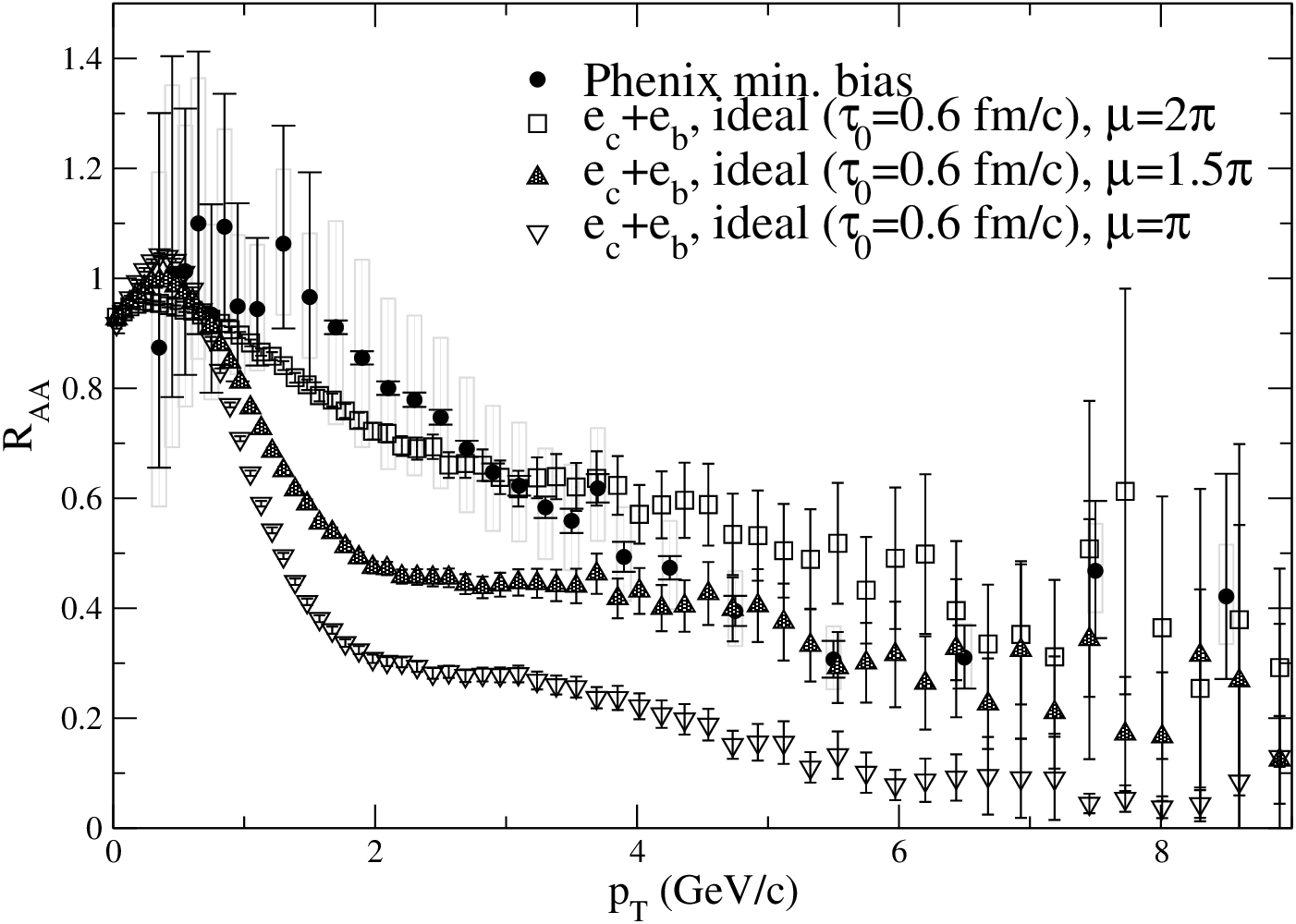}
\includegraphics[clip,width=0.495\textwidth]{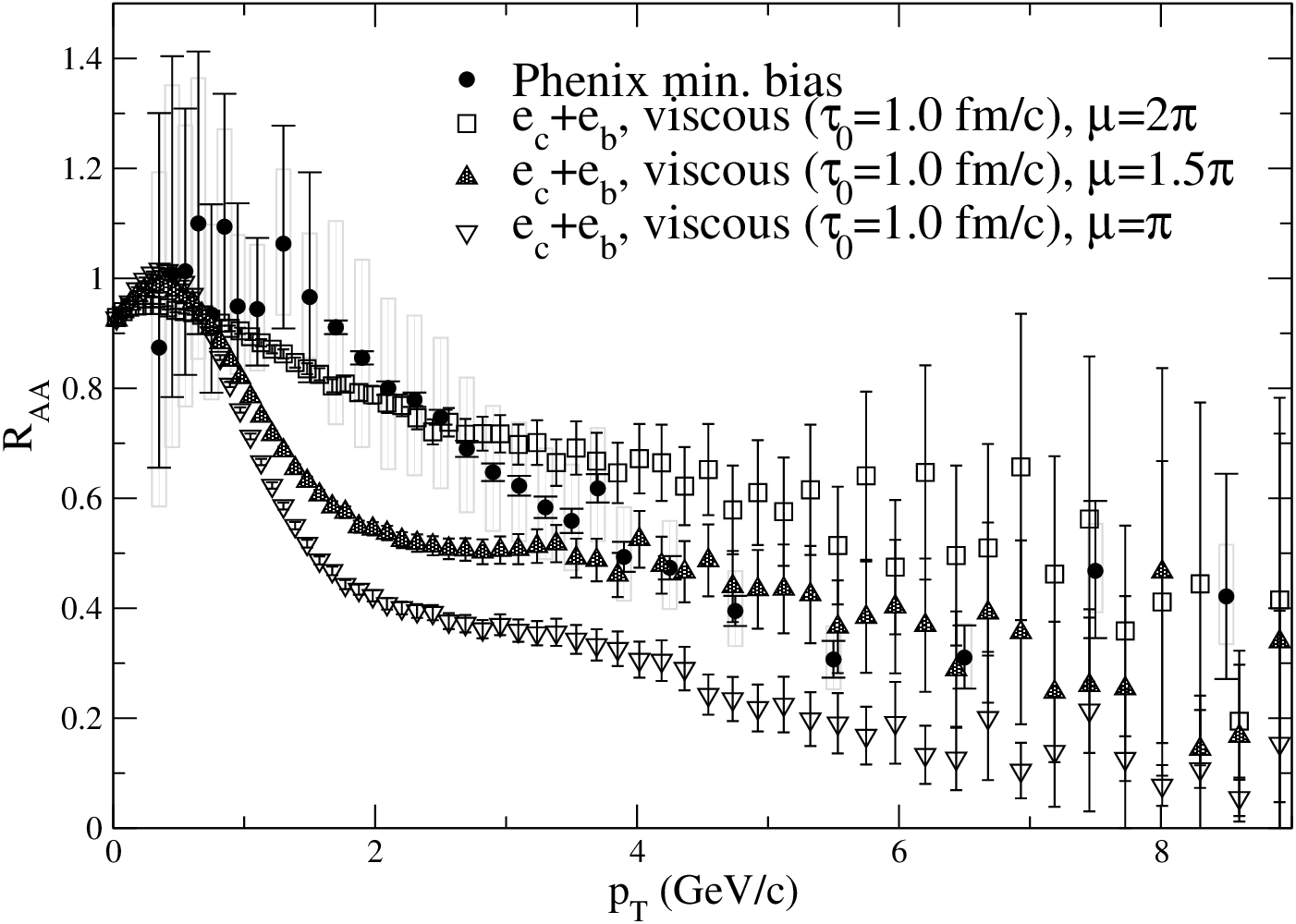}
\caption{The electron $R_{AA}$ for minimum bias Au-Au collisions at $\sqrt{s}_{\rm NN}\!=\!200$ GeV. The sensitivity to the scale $\mu\!=\!\pi T-2\pi T$ is displayed. The high-momentum region ($p_T\!\simge\!4$ GeV/c) is better reproduced with the intermediate coupling. Ideal (left panel) and viscous (right panel) hydro scenarios provide similar results.} 
\label{fig:RAA}
\end{center}
\end{figure}
We first consider the nuclear modification factor $R_{AA}(p_T)\!\equiv\!(dN/dp_T)^{AA}/\langle N_{\rm coll}\rangle(dN/dp_T)^{pp}$ of the heavy-flavor spectra. In Fig.~\ref{fig:RAA_system} we display the effects of fragmentation and semi-leptonic decays on the charm $R_{AA}$, both leading to an additional quenching of the spectrum. We also show the separate contribution of the electrons from charm and bottom. In Fig.~\ref{fig:RAA} the $R_{AA}$ of non-photonic electrons measured by Phenix is compared with our findings for different values of the coupling. The high-momentum region ($p_T\!\simge\! 4$ GeV/c) is better reproduced evaluating $\alpha_s$ at the scale $1.5\pi T$. The dependence on the hydro scenario (ideal/viscous) turns out to be small.
\begin{figure}
\begin{center}
\includegraphics[clip,width=0.495\textwidth]{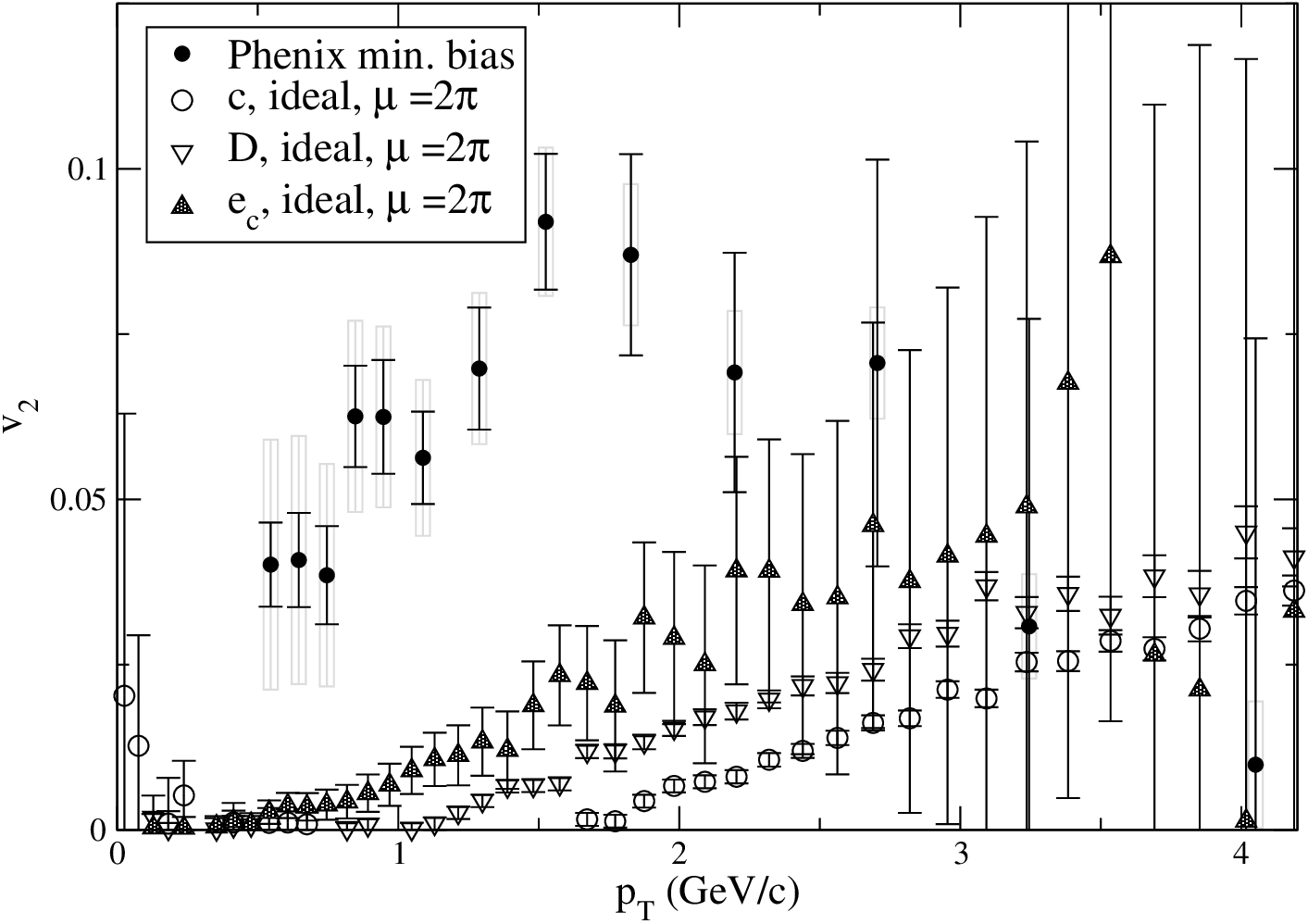}
\includegraphics[clip,width=0.495\textwidth]{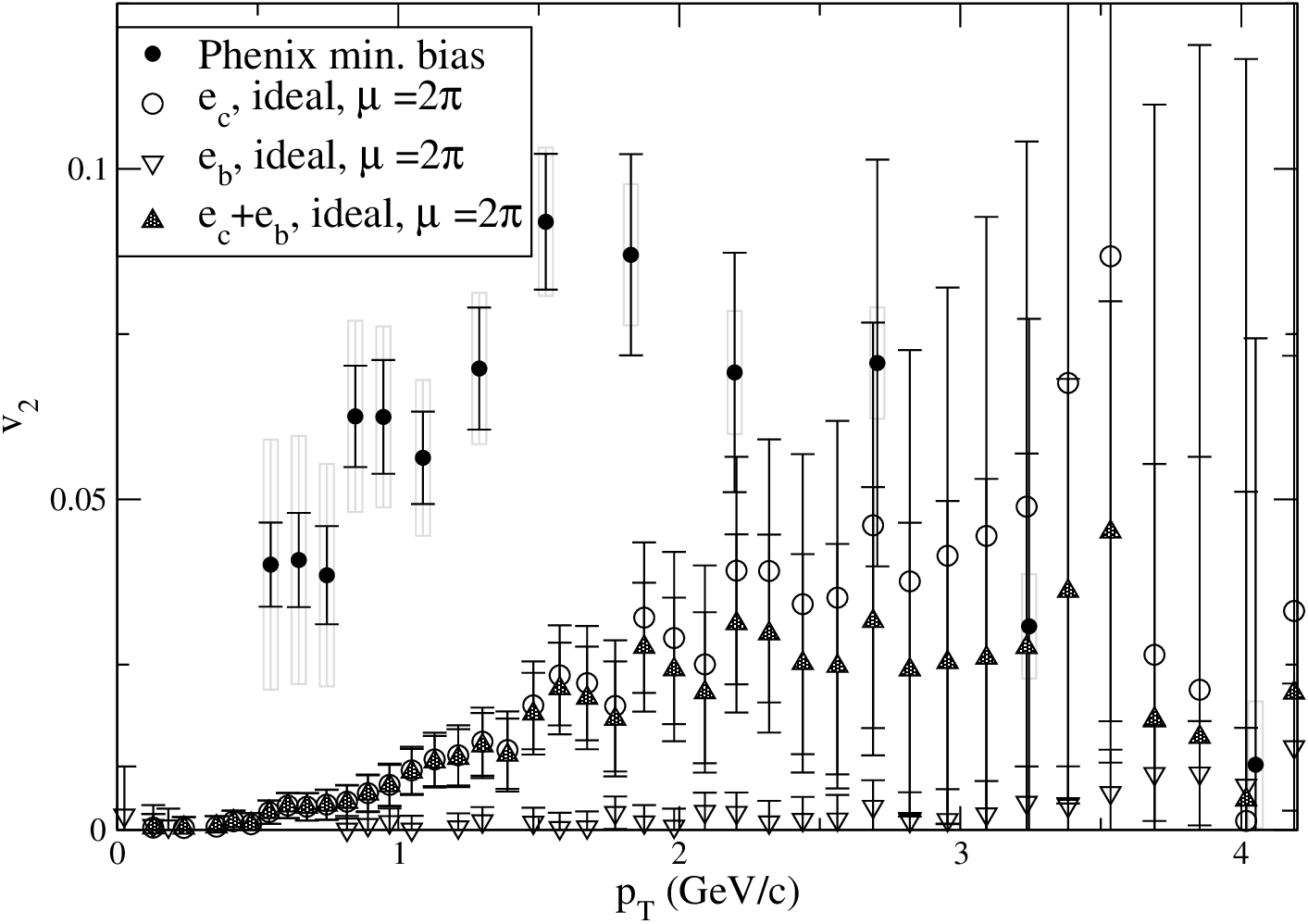}
\caption{Left panel: the effect of fragmentation and semi-leptonic decays on the elliptic-flow of charm, resulting in a shift to lower values of $p_T$. Right panel: the separate contribution of charm and bottom to the single-electron $v_2$. The bottom does not display any flow.} 
\label{fig:v2_system}
\end{center}
\end{figure}
\begin{figure}
\begin{center}
\includegraphics[clip,width=0.495\textwidth]{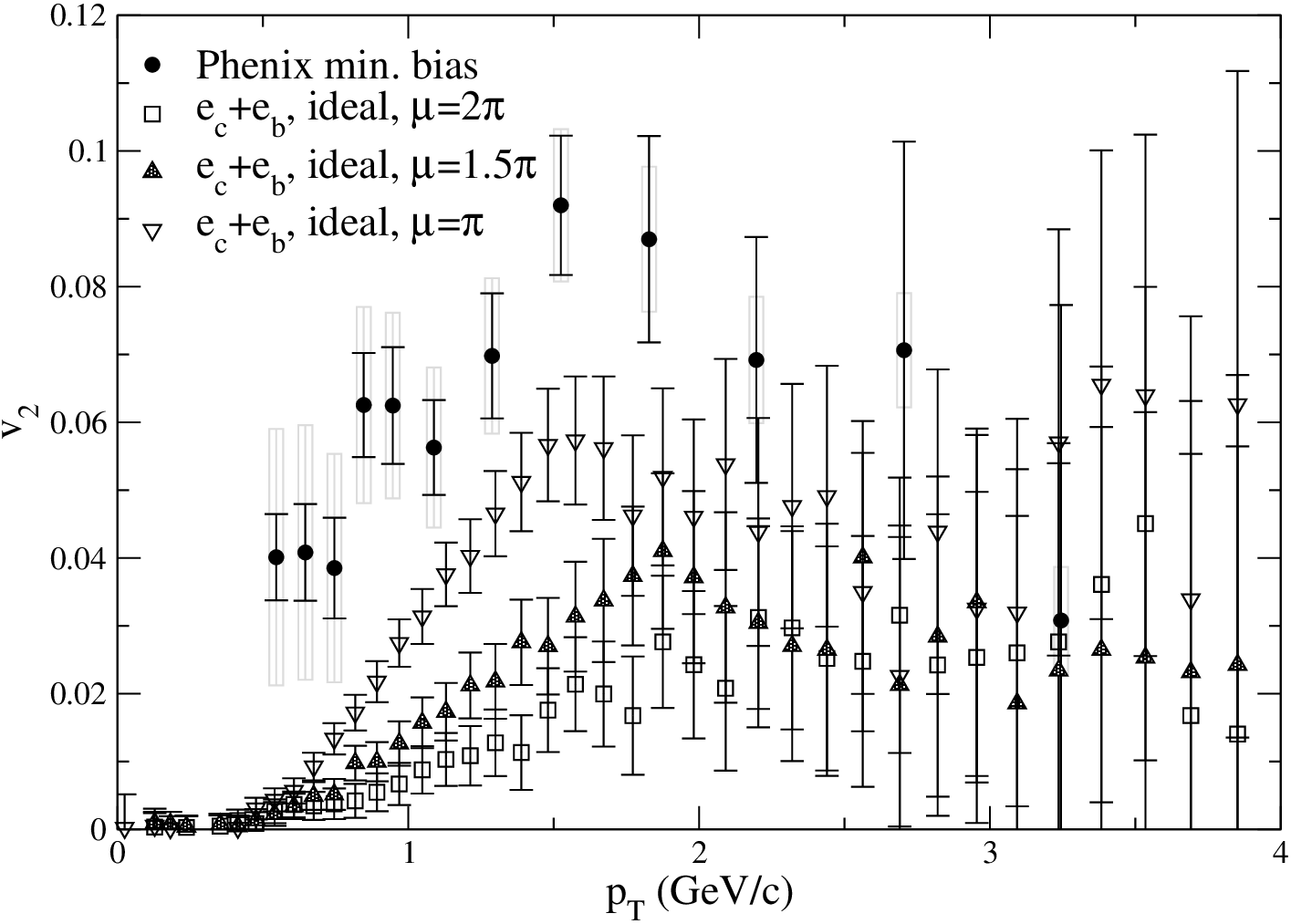}
\includegraphics[clip,width=0.495\textwidth]{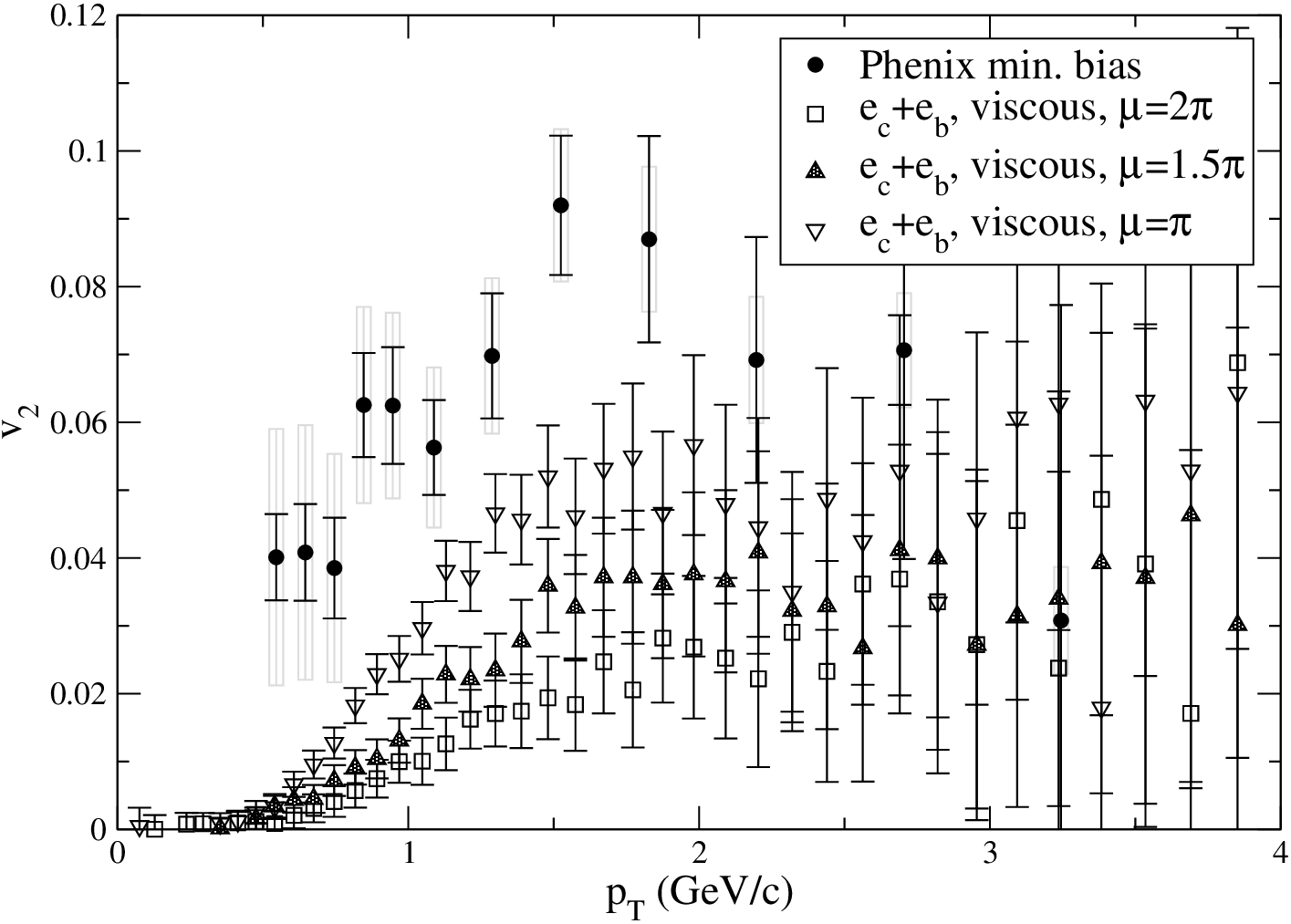}
\caption{The elliptic-flow coefficient $v_2$ of non-photonic electrons for minimum bias Au-Au collisions at $\sqrt{s}_{\rm NN}\!=\!200$ GeV. The sensitivity to the scale $\mu\!=\!\pi T-2\pi T$ is displayed. The data would favor the strongest value of the coupling, which however would provide a too small value of $R_{AA}$. Ideal (left panel) and viscous (right panel) hydro scenarios provide again similar results.} 
\label{fig:v2}
\end{center}
\end{figure}
In Figs.~\ref{fig:v2_system} and~\ref{fig:v2} the same kind of analysis is performed for the elliptic-flow coefficient $v_2(p_T)\!\equiv\!\langle\cos(2\phi)\rangle_{p_T}$, with $\phi$ measured with respect to the reaction plane. Our results for the electron $v_2$ underestimate the experimental data. The same occurs for $R_{AA}$ at moderate $p_T$. Part of this disagreement could arise from ignoring coalescence as a mechanism of hadronization, as can be inferred from the analysis of its effect performed in Ref.~\cite{rapp}. A possible further discrepancy would require a better evaluation of the transport coefficients.


\begin{thebibliography}{30}
\bibitem{lange_hot} W.M. Alberico \emph{et al.}, arXiv:1007.4170 [hep-ph]. 
\bibitem{hira} Y. Akamatsu, T. Hatsuda, T. Hirano,
               Phys. Rev. C {\bf 79} 054907 (2009).
\bibitem{rapp} H. van Hees, V. Greco and R. Rapp,
               Phys. Rev. C {\bf 73}, 034913 (2006).
\bibitem{tea} G.D. Moore, D. Teaney,
              Phys. Rev. C {\bf 71}, 064904 (2005).
\bibitem{aic} P.B. Gossiaux and J. Aichelin,
              Phys. Rev. C {\bf 78}, 014904 (2008).
\bibitem{kolb1} P.F. Kolb, J. Sollfrank and U. Heinz,
                Phys. Rev. C {\bf 62}, 054909 (2000).
\bibitem{rom1} P. Romatschke and U.Romatschke,
               Phys. Rev. Lett. {\bf 99}, 172301 (2007). 
\bibitem{rom2} M. Luzum and P. Romatschke,
               Phys. Rev. C {\bf 78}, 034915 (2008). 
\bibitem{lange} A. Beraudo, A. De Pace, W.M. Alberico and A. Molinari,
                Nucl. Phys. A 831, 59 (2009).
\bibitem{pei} S. Peign\'e and  A. Peshier,
               Phys. Rev. D {\bf 77}, 114017 (2008).
\bibitem{com} B.L. Combridge,
              Nucl. Phys. B {\bf 151}, 429 (1979).
\bibitem{POWHEG} S. Frixione, P. Nason and G. Ridolfi,
                 JHEP 0709, 126 (2007).
\bibitem{EPS09} K.J. Eskola, H. Paukkunen and C.A. Salgado,
              JHEP 0904, 065  (2009).
\bibitem{peter} C. Peterson, D. Schlatter, I. Schmitt and P.M. Zerwas,
                Phys. Rev. D {\bf 27}, 105 (1983).
\bibitem{zeus} http://www-zeus.desy.de/zeus\_papers/ZEUS\_PAPERS/DESY-05-147.ps .
\bibitem{pdg} http://www.slac.stanford.edu/xorg/hfag/osc/PDG\_2009/\#FRAC .
\bibitem{Pythia} T. Sjostrand, S. Mrenna and P.Z. Skands,
                 JHEP 0605, 026 (2006).
\bibitem{pdg09} C. Amsler \emph{et al}. (PDG), Phys. Lett. B667, 1 (2008)
                and 2009 partial update.
\bibitem{Phenix} A. Adare \emph{et al.} (PHENIX Collaboration),
              Phys. Rev. Lett. {\bf 98},  172301 (2007).
\end{thebibliography}
\end{document}